\documentclass[preprint2]{aastex}

\usepackage{color}

\newcommand{\Msol}{M_\odot}
\newcommand{\kms}{\rm{km~s^{-1}}}
\newcommand{\Mpc}{\rm Mpc}
\newcommand{\Gyr}{\rm Gyr}
\newcommand{\Gyrs}{\rm Gyrs}
\newcommand{\Mpch}{\rm Mpc \, \it h^{-1}}
\newcommand{\kpch}{\rm Kpc \, \it h^{-1}}
\newcommand{\kpc}{\rm Kpc}

\shorttitle{Tidal stripping of globular clusters}
\shortauthors{Ramos et al.}

\begin{document}


\title{Tidal stripping of globular clusters in a simulated galaxy cluster}


\author{F. Ramos, V. Coenda \altaffilmark{1}, H. Muriel \altaffilmark{1}, and M. Abadi \altaffilmark{1}}
\affil{Instituto de Astronom\'{\i}a Te\'orica y Experimental, CONICET-UNC, Laprida 922, C\'ordoba, Argentina}


\altaffiltext{1}{Observatorio Astron\'omico de C\'ordoba, UNC, Laprida 854, C\'ordoba, Argentina}


\begin{abstract}
Using a cosmological N-body numerical simulation of the formation of a galaxy cluster- sized halo, we analyze 
the temporal evolution of its globular cluster population. We follow the dynamical evolution of 38 galactic dark matter halos 
orbiting in a galaxy cluster that at redshift $z=0$ has a virial mass of $1.71 \times 10 ^{14} \Msol h^{-1}$. 
In order to mimic both "blue" and "red" populations of globular clusters, for each galactic halo we select 
two different sets of particles at high redshift ($z \approx 1$), constrained by the condition that, at redshift $z=0$, their 
average radial density profiles are similar to the observed profiles. As expected, the general galaxy cluster tidal field 
removes a significant fraction of the globular cluster populations to feed 
the intracluster population. On average, halos lost approximately 16\% and 29\% of 
their initial red and blue globular cluster populations, respectively. Our results suggest that these fractions strongly depend on  
the orbital trajectory of the galactic halo, specifically on the number of orbits and on the minimum 
pericentric distance to the galaxy cluster center that the halo has had. At a given time, these fractions also depend 
on the current clustercentric distance, just as observations show that the specific frequency
of globular clusters $S_N$ depends on their clustercentric distance.
\end{abstract}


\keywords{galaxies: clusters: general --- galaxies: star clusters: general --- methods: numerical}



\section{Introduction}
\label{sec:intro}

Galaxy clusters are extreme environments where thousands of galaxies 
orbit inside a massive system of $ \sim  10^{14} \Msol$, a few $\Mpc$ size,
and with a typical velocity dispersion of $ \sim 1000\, \kms$.
In these regions, galaxies evolve under the influence of the strong general 
tidal field generated by the cluster’s gravitational potential and also by 
individual mutual gravitational interactions with other galaxies 
\citep[e.g.][]{Merritt:1983,Merritt:1984,Malumut:1984}. 

From the theoretical point of view, it is expected that the global tidal 
field in a massive galaxy cluster will be strong enough to truncate 
dark matter halos of individual galaxies \citep{Merritt:1983}. 
\citet{Diemand:2007} found that the 
removal of the dark matter halo happens from the outside in,
i.e. the outer particles of a halo are less bound and thus more prone
to be removed.
Although the baryonic material is much 
more concentrated than the dark matter, it can also suffer the effects of the 
galaxy cluster tidal field. 
Using N-body/hydrodynamic simulations, \citet{Limousin:2009} investigated 
the effect of tidal stripping on galaxy clusters, tracking 
both the dark matter halo and the stellar component, 
and found that the dark matter component is preferentially stripped, 
while the stellar component is less affected by tidal forces.

Globular cluster systems are among the most extended galactic stellar 
components; they can reach out well beyond the detectable stellar halo and are also prone to  
being tidally stripped. Since dark matter density in central regions of galaxy
clusters is expected to be higher than in their outskirts,
this could result in more compact halos in these central regions 
\citep{Bull:2001}. A similar effect could be expected for globular clusters, lowering their 
specific frequency $S_N$ (number per luminosity unit) in the central regions of galaxy clusters 
compared to the outskirts. \citet{Forbes:1997} found evidence of this effect in four galaxies in 
the central region of the Fornax cluster with a marginal dependence of the specific 
frequency on the clustercentric distance. 
\citet{Coenda:2009} searched for similar evidence using data from the
ACS Virgo Cluster Survey on board the Hubble Space Telescope.
These authors analyzed the $S_N$ in 13 elliptical galaxies and found that $S_N$ 
increases as a function of both projected and 3D Virgo clustercentric distance,
while the projected number density of background globular clusters 
decreases. 
These results are interpreted as direct evidence of tidal 
stripping of globular clusters due to the gravitational potential of the galaxy cluster. 
In a similar sense, \citet{Alamo-martinez:2013}
studied the globular cluster system in the center of the massive galaxy cluster 
Abell 1689 and estimated that $80,000$ out of a population of $162,000$ are 
possibly part of the intracluster medium  \citep[see also][]{Webb:2013}. 
Evidence of tidal stripping of globular clusters has also been 
found in less dense environments \citep[see for instance][]{Blom:2014}.
\citet{Hudson:2014} use galaxy halo masses based on weak 
lensing to compute the ratio between the total mass in globular clusters and the halo mass. They 
found that this ratio is constant with an intrinsic scatter of $0.2$ dex and suggested that 
some of this scatter is due to differing degrees of tidal stripping of the globular cluster systems  
between central and satellite galaxies.

In the last two decades or so, growing evidence has been collected of a
bimodal color distribution of globular clusters 
\citep{Gebhardt:1999,Kundu:2001,Brodie:2006}.
This is usually interpreted as the consequence of differences in 
metallicity between two co-existing globular cluster populations:
one red and metal-rich $[Fe/H]\sim -0.5$ and one blue and metal-poor $[Fe/H]\sim -1.5$ \citep{Brodie:2006}.
\citet{Peng:2006} found bimodal color distributions for globular clusters 
in approximately 100 early-type galaxies of the Virgo galaxy cluster. 
They found that, on average, red globular clusters account for 15\% of the total population in faint ($M_B \sim -15.7$) 
galaxies and up to 60\% in bright ($M_B \sim -21.5$) galaxies. 
Using the same observational data, \citet{Coenda:2009} fitted power-laws to the projected number density profiles, 
finding typical logarithmic slope values of $\alpha=-2.4 \pm 0.2$ and $\alpha=-1.7 \pm 0.1$ for red and blue populations, 
respectively. Although with a large scatter, other authors \citep[see for example][]{Larsen:2001,Bassino:2008,Richtler:2012,
Abrusco:2014} agree that blue GCs usually show a shallower distribution than
that corresponding to red GCs in the whole range of radii. 
Since the blue population is more extended than the red, 
it is expected that it will be more prone to be tidally disrupted than the red. 
Indeed, $S_N$ dependence on clustercentric distance reported by 
\citet{Coenda:2009} has been detected only for blue 
but not for red globular clusters. \citet{Kartha:2014} compiled a sample of 40 galaxies
and found that the relative fraction of blue to red globular clusters 
decreases with increasing galaxy environmental density for 
lenticular galaxies. They propose that the outer blue globular cluster population may be stripped 
away, giving a lower ratio of blue to red.

From the theoretical point of view, an early work of \citet{Muzzio:1984} analyzed the
swapping of globular clusters in Virgo-like cluster of galaxies using ad-hoc N-body 
numerical simulations. 
These authors show that (smaller) galaxies in clusters can lose up to 30\% of their 
globular cluster population, with half of this fraction ending up in the intracluster 
medium and the other half captured by other (larger) galaxies.
\citet{B&Y:2006} selected globular cluster particles in 
cosmological N-body numerical simulations  
and found that intracluster globular clusters can 
contribute up to $40\%$ of the total globular cluster population in 
galaxy clusters of mass $1.0$ to $6.5 \times 10^{14} \Msol$. 
More recently, \citet{Smith:2013} studied the impact of harassment on the 
dynamics of globular clusters in early-type dwarf galaxies of clusters. 
These authors modeled galaxies that evolved for $2.5\, \Gyr$ within a potential field 
that mimics the effects of harassment in clusters, finding 
that the dynamical behavior of the globular clusters strongly 
depends on the fraction of bound dark matter remaining in the galaxy. 
They conclude that when 85\% of the initial dark matter halo mass 
has been removed, globular clusters begin to be stripped.
  
Using N-body numerical simulations in the framework of the $\Lambda$ Cold 
Dark Matter cosmological model, we explore the tidal removal of globular 
clusters in a Virgo-like galaxy clusters, focusing on the differences between red and blue 
populations. 
In Section \ref{sec:nsim} we present the numerical simulation used in this work.
In Section \ref{sec:model} we describe the method that we apply in order
to select dark matter particles from simulated galactic halos that mimic 
their globular cluster population.
In Section \ref{sec:results} we present the main results about the removal
of red and blue globular cluster populations. 
Finally, in Section \ref{sec:conclusions} we discuss the main results and conclusions.

\section{The numerical simulation}
\label{sec:nsim}
 
\begin{figure}
\includegraphics[width=7cm]{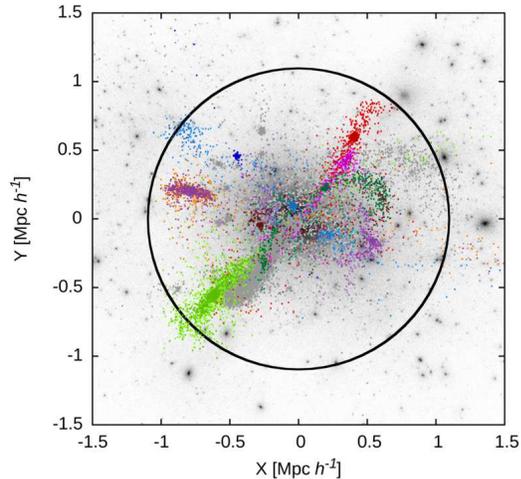}
  \caption{Projected spatial distribution of dark matter particles (small gray dots) centered in the simulated galaxy 
cluster at redshift $z=0$. The solid circle shows the system virial radius. 
Different colors show dark matter particles highlighted as blue globular clusters from nine halos, 
following the method described in Section \ref{sec:model}.}
 \label{fig:paint}
\end{figure}

We address these issues using a dark matter-only cosmological numerical 
simulation that follows the formation of a galaxy cluster-sized halo. 
This simulation was presented in detail by \citet{Ludlow:2010} to which we refer 
the interested reader for further technical details. 
In brief, a region destined to form a galaxy cluster at redshift $z=0$ is 
identified in a larger cosmological box of $100 \,\Mpch$ comoving on a side. 
Then, this region is resimulated at a higher resolution using a zoom-in 
technique \citep{Klypin:2001} while the remainder of the box is resampled at a lower resolution.
The high-resolution region has $1.44\times 10^7$ dark matter particles filling 
an amoeba-shaped volume of $~1.39\, \Mpc^3 h^{-3}$ at a starting redshift of $z=19$. 
The original simulation assumes a $\Lambda$ Cold Dark Matter cosmology with the following 
parameters: $H_0 = 73\, \kms \Mpc^{-1}$ (i.e. $h = 0.73$), $\Omega_0 = 0.25$ and $\Omega_{\Lambda}=0.75$, 
Both simulations were performed with the GADGET2 code \citep{Springel:2005}.
High resolution dark matter particle masses of the resimulation are
$5.4 \times 10^7 \Msol h^{-1}$ and the gravitational softening is $1.5\,\kpch$ comoving.
At redshift $z = 0$, the simulated galaxy cluster has a virial
radius of $R_{vir} = 1.1 \,\Mpch$, which corresponds to a virial mass
$M_{vir} = 1.71 \times 10^{14} \Msol\,h^{-1}$. These values are comparable to those estimated 
for the Virgo Cluster \citep[see][]{Karachentsev:2014}.
Within the virial radius (i.e. radius where the average mass density is 200 
times the critical mass density of the universe) at 
redshift $z = 0$ there are $3.16\times 10^6$ dark matter particles. This halo 
corresponds to the halo labeled "h14" in Table 1 of \citet{Ludlow:2010} which is isolated and virialized at z=0.
In order to select the main cluster-sized halos and their galactic subhalos,
we use the SUBFIND algorithm \citep{Springel:2001}
which identifies substructures inside friends of friends groups.
We run SUBFIND on all 100 snapshots, equally spaced in $\log(a)$, from redshift $z=19$ to $z=0$ to
follow the dynamical evolution of their globular cluster populations.
In Fig. \ref{fig:paint} we show the projected spatial distribution of
dark matter particles (small gray dots) centered in the simulated galaxy 
cluster at redshift $z=0$, where the solid circle shows the system virial
radius. Using different colors we highlight dark matter particles selected as nine
different globular cluster systems, following the method described in the
next section. All the remaining globular cluster systems are depicted as gray dots.

\section{The model}
\label{sec:model}

\subsection{Globular cluster population}

Using SUBFIND \citep{Springel:2001} we identified the 158 most massive dark matter 
halos that at redshift $z=0$ are located inside the virial radius of the 
galaxy cluster. At z=0, the less massive halo is resolved by at least 200 particles, which corresponds
to $M \sim 1 \times 10^{10} M_{\odot}h^{-1} $, while the most massive one has about 4000 particles, 
which corresponds to $M \sim 2 \times 10^{11} M_{\odot}h^{-1} $.
We follow the temporal evolution of each halo, computing its mass, density profile, orbit and 
accretion time i.e. the time when it enters the galaxy cluster virial radius for the first time.
In order to perform reliable statistics, we limited our sample to those halos that, 
at selection time, were resolved by at least 700 particles.
In order to maximize the amount of time that halos are orbiting the galaxy cluster,
we exclude from our analysis those halos that have late accretion time 
($t > 10\, \Gyr$). We have also excluded 19 halos that at the selection time (or even before) had a near neighbor halo. 
Our final sample has 38 halos.

\begin{figure}
\includegraphics[width=7cm,angle=270]{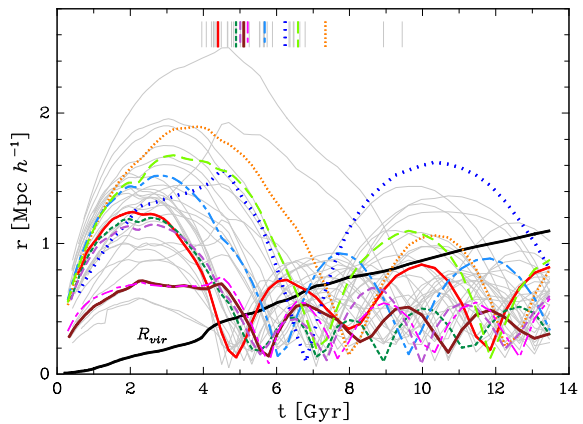}
  \caption{Clustercentric distances as a function of time for 38 halos (gray lines). Colors correspond to the nine halos 
  shown in colors in Fig. \ref{fig:paint}. The black solid line is the virial radius of the cluster. 
   Vertical lines show the time that each halo enters the cluster virial radius.}
\label{fig:orbitas}
\end{figure}

In Fig. \ref{fig:orbitas} we show the clustercentric distance as a function 
of time for these 38 halos (gray lines), highlighting the nine halos also 
highlighted with the same color-coding in Fig. \ref{fig:paint}. 
In order to minimize halos with possible tidal distorsions, we 
checked each 
halo during the infall process. We noticed that at accretion time many halos are already distorted 
by the cluster’s gravitational potential. However, two snapshots before that, halos show no evidence 
of such distortion. Consequently, we adopt this time as the snapshot to select 
globular clusters.
Selection time was, on average, $0.5 \Gyr$ before the accretion time 
(shown as vertical lines in Fig.\ref{fig:orbitas}).  
We computed the mass density profile of halos at this time, 
fitting a NFW profile \citep{NFW:1996} expressed as:
\begin{equation}
\rho_{NFW}(r)=\frac{\rho^0_{NFW}}{(r/r_{NFW})(1+r/r_{NFW})^2} 
\end{equation}
where $\rho^0_{NFW}$ and $r_{NFW}$ are the dark matter halo’s characteristic density and scale length, respectively. 
Typical accretion cosmic times range between $4.0$ and $9.3 \,\Gyrs$ with a median of $5.2 \,\Gyrs$.
Scale lengths range between $7.5$ to $87.8\, \kpch$ with
a median of $10.5 \,\kpch$ while virial masses range between $5.3 \times 10^{10} $ and $1.7 \times 10^{13} M_{\odot}h^{-1}$ with 
median $1.7 \times 10^{11} M_{\odot}h^{-1}$. In order to build up globular cluster populations of each halo, we selected a subsample of 
their dark matter particles, using the method outlined by \citet{Bullock:2005} and \citet{Pena:2008}. 
In this method, a subset of particles is chosen to follow a given spatial mass 
density profile $\rho(r)$ following its corresponding energy distribution function:
\begin{equation}
f(\epsilon)=\frac{1}{8\pi}
\Big[\int_0^\epsilon\frac{d^2\rho}{d\psi^2}\frac{d\psi}{\sqrt{\epsilon-\psi}}+\frac{1}{\sqrt{\epsilon}}\Big(\frac{d\rho}{d\psi}\Big)_{\psi=0}\Big]
\label{ec:fdist}
\end{equation}
where $\psi$ is the relative gravitational potential and $\epsilon$ is the relative energy \citep[see][]{BT:1987}. 

In order to construct blue and red globular cluster populations for each halo
at accretion time
we generate two analytic \citet{HQT:1990} density profiles 

\begin{equation}
\rho_{H}(r)=\frac{\rho^0_{H}}{(r/r_{H})(1+r/r_{H})^3}
\end{equation}

where $r_H$ is the scale length and $\rho_H^0$ is the characteristic density. 
We scaled $r_H= \gamma r_{NFW}$ adopting $\gamma=3$ for blue and 
$\gamma=0.5$ for red globular cluster populations. 
With this choice for $\gamma$ at accretion time, the projected density profiles at redshift 
$z=0$ are approximately power laws with slopes similar to those observed for blue and red populations (see Section \ref{sec:intro}).
At accretion time, we apply Equation (\ref{ec:fdist}) to the \citet{HQT:1990} and \citet{NFW:1996} profiles in order to
construct their corresponding energy distributions $f_H(\epsilon)$ and $f_{NFW}(\epsilon)$, respectively.

\begin{figure}
\includegraphics[width=7cm]{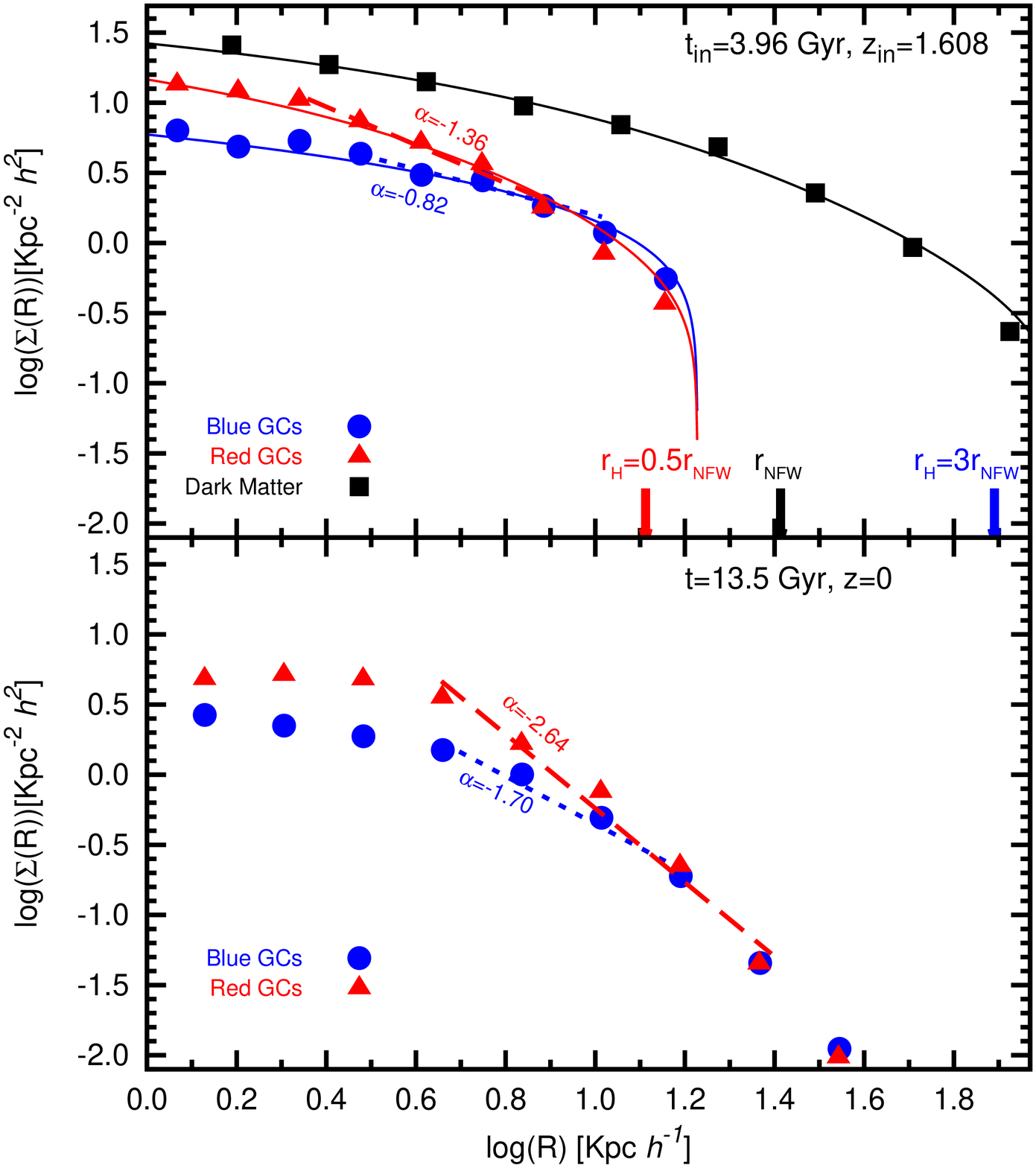}
  \caption{Projected radial density profile of one selected halo for dark matter particles (solid black squares), for blue (blue circles) and red 
  (red triangles) globular clusters at redshift $z_{in}$ (\emph{top panel}) in which the halo enters the cluster's virial radius, and at 
  $z=0$ (\emph{bottom panel}). \emph{Top panel}:  The dark halo is modeled as a NFW profile (solid black line), while blue and red 
  globular clusters are modeled as Hernquist profiles (solid blue and red lines, respectively). Vertical arrows show the Hernquist scale 
length $r_H$ for blue ($3r_{NFW}$) and red ($0.5r_{NFW}$) globular cluster populations,
  and the NFW scale length ($r_{NFW}$) for the dark matter halo. In \emph{both panels} we show a power law fit to the data. Short dashed lines correspond to 
  blue globular clusters and long dashed lines to red globular cluster populations. 
}
 \label{fig:profiles}
\end{figure}

Then, we randomly select a fraction $f_H(\epsilon)/f_{NFW}(\epsilon)$ of dark matter particles
in bins of specific energy $\epsilon + \Delta \epsilon$.
Following \citet{B&Y:2006} we assume that the baryonic component is more concentrated than the dark matter halo, so
we truncate the selection at a cut off radius of $r_{CutOff}=r_{50}/3$, where $r_{50}$ is the half-mass radius
of the dark matter halo. As an example, in Fig. \ref{fig:profiles}, we show the projected density profiles 
$\Sigma(R)$ of one selected halo as a function of the projected distance $R$, for the dark 
matter halo (black squares) and for the selected particles chosen to 
represent the blue (blue circles) and red (red triangles) globular cluster populations.
In the \emph{top panel} we show the profile at its accretion time ($t = 4.9\,\Gyrs, z_{in}=1.25$) and
in the \emph{bottom panel} at the present time ($t=13.7\, \Gyrs, z = 0$).
This dark matter halo corresponds to that shown in light green in Fig. \ref{fig:paint}, \ref{fig:orbitas} 
and \ref{fig:fraction}. The solid black line corresponds to a \citet{NFW:1996} density profile
with concentration $c=5.9$ and virial mass $M_{vir}=2.6 \times 10^{12} M_{\odot}h^{-1}$.
Blue and red solid lines are not fits to the points, but rather show the 
analytic \citet{HQT:1990} profiles chosen to represent both globular cluster populations.
Dashed lines show power law fits to the points where we indicate the
corresponding slope $\alpha$ for each globular cluster population at
accretion time.
We follow the temporal evolution of these blue and red populations in order
to compare their properties with the corresponding populations observed
in galaxy clusters and we present the main results in the next section.

\section{Results}
\label{sec:results}

\subsection{Density profile of globular clusters}

\begin{figure}
         \includegraphics[width=7cm]{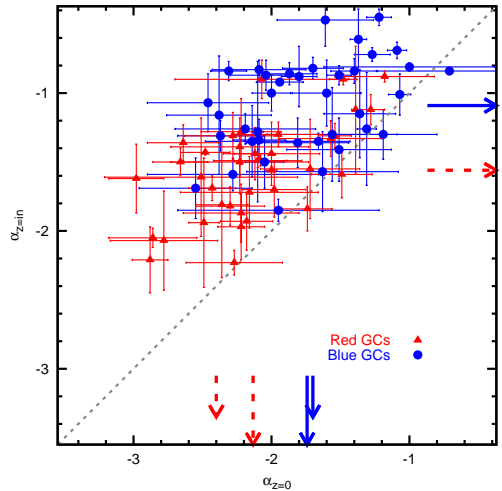}
	\caption{Slopes of the projected density profiles of red and blue globular clusters at the redshift 
	$z_{in}$ in which they enter the cluster's virial radius, versus the corresponding slopes at $z=0$. 
	Symbol and color types as in Fig. \ref{fig:profiles}. Horizontal and long vertical arrows 
correspond to the median values of slopes for blue (solid blue line) and red (dashed red line) globular cluster populations.
Short vertical arrows indicate observed slope values for red and blue CGs. 
Gray dashed line shows the identity $\alpha_{z_0}=\alpha_{z_{in}}$.}
	\label{fig:slopes}
\end{figure}

As can be seen in Fig. \ref{fig:slopes}, the density profile of globular clusters is steeper at redshift zero than at the time
of selection. We analyzed the evolution of these profiles as a function of redshift and found that the profiles
become steeper immediately after the halos enter the cluster. After that, slopes remain nearly constant 
independently of the evolution of the associated globular clusters. This indicates that, even in those cases where 
an important fraction of globular clusters is removed, the remaining bound objects are redistributed in order to preserve a 
certain density profile. This is in agreement with  \citet{Coenda:2009}, who found a lack of dependence 
between the slopes of the density profiles of globular clusters and the clustercentric distances in the Virgo cluster. 
Nevertheless, \citet{Bekki:2003}, using a simple numerical model, found that the radial number 
density of the globular cluster system becomes steeper after the stripping of globular clusters. 

\subsection{Globular cluster fraction evolution}

In order to follow the temporal evolution of globular clusters orbiting 
the galaxy cluster, for each halo, we estimate the fraction
that is associated with its own main halo at any given redshift.
We assume that a globular cluster is associated with its main halo if
its distance to the halo center is less than the most distant dark 
matter halo particle identified by SUBFIND \citep{Springel:2001} as part of this halo. 
In order to compute this fraction, at any given redshift $z$, we compute the 
ratio between the globular cluster number that we have selected at accretion 
redshift $z_{in}$ and the globular cluster number contained inside a sphere 
of radius equal to the most distant dark matter particle identified by SUBFIND.
Although this selection criterion may result in the inclusion of globular 
clusters that are not dynamically linked to the halo, this criterion mimics what 
is usually done with the observational data, where globular clusters are selected 
up to a given radius. Moreover, we have found that our results are fairly robust if 
we adopt a binding energy instead of our distance criteria; average differences in the 
fractions plotted in figures 5, 6, 7 and 8 are always less than 7\%.

\begin{figure}
\includegraphics[width=7cm]{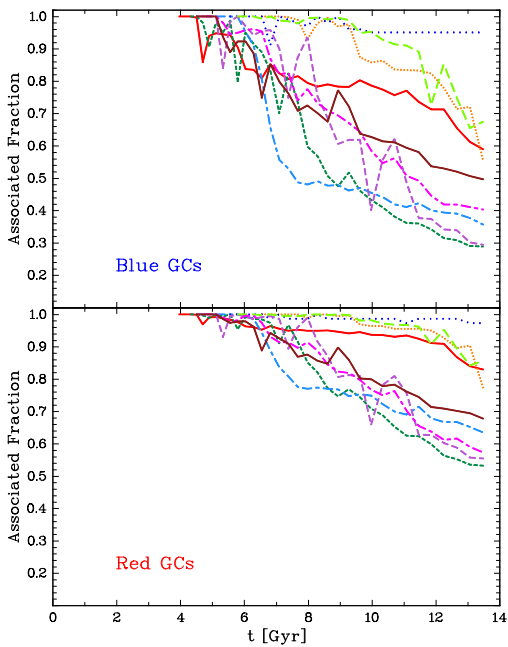}
  \caption{Associated fraction of globular clusters as a function of time for blue (\emph{top panel}) and red 
  (\emph{bottom panel}) globular clusters
  for the same nine halos highlighted in Fig. \ref{fig:paint}. Color-coding as in Fig.\ref{fig:paint}.}
 \label{fig:fraction}
\end{figure}

Fig. \ref{fig:fraction} shows the fraction of globular clusters associated 
with its main halo as a function of time for both blue (\emph{top panel}) and red 
 (\emph{bottom panel}) globular clusters (color coding for this figure is 
the same as that of Fig. \ref{fig:paint}).
The general trend is that at accretion time $t_{in}$ this fraction is, by definition 
$f(z_{in})=1$ and decays to $f(z=0) \sim 0.6(0.75)$ at redshift $z=0$ which means that
about 40\%(25\%) of blue(red) globular clusters are removed from its halo to 
become part of the intracluster population.
The steeper slopes in the upper panel of Fig. \ref{fig:fraction} clearly 
show that blue globular clusters are more easily removed than red, as
expected  
due to the shallower spatial distribution of the blue population (see Fig. \ref{fig:slopes}). 
Some temporal oscillations are noticeable in the fraction of globular clusters
associated with its halo (see for example the purple dashed line).
Fig. \ref{fig:orbitas} and \ref{fig:fraction} show that the local minima 
correspond to the successive passages of halos through 
the center of the galaxy cluster. However, it should be noticed that 
globular clusters are continuously removed throughout the lifetime of halos not
only due to pericentric passages.
On the other hand, local maxima appear when their orbits intersect a stream of previously removed 
globular clusters. Also, 
we followed the fate of the stripped globular clusters and found that none of them are recaptured by other halos.

\subsection{Globular cluster fraction and orbital trajectory}

The strength of the tidal field generated by the gravitational 
potential of the galaxy cluster
depends on the relative position of halos with respect to the cluster center. 
Thus, a correlation is expected between the fraction of globular cluster that
remains associated with its halo and its orbital parameters. 
Consequently, we searched for correlations between the fraction of globular 
clusters associated with its halo and the number of orbital revolutions completed 
by the halo from accretion time $t_{in}$ to redshift $z=0$. 

\begin{figure}
\includegraphics[width=7cm]{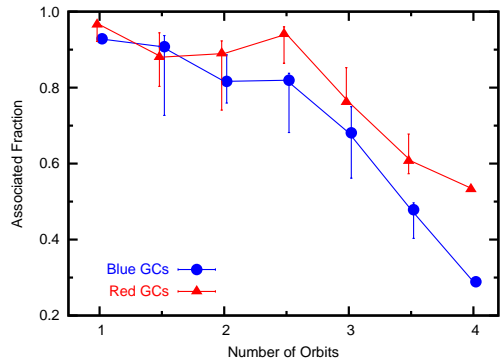}
  \caption{Associated fraction of globular clusters as a function of the number of orbits at $z=0$ for blue and red globular clusters.
  Points correspond to median values and the vertical error bars are computed using the bootstrap resampling technique. Symbol and 
  color types as in Fig. \ref{fig:profiles}.}
\label{fig:fraccion_orb}
\end{figure}

\begin{figure}
\includegraphics[width=7cm]{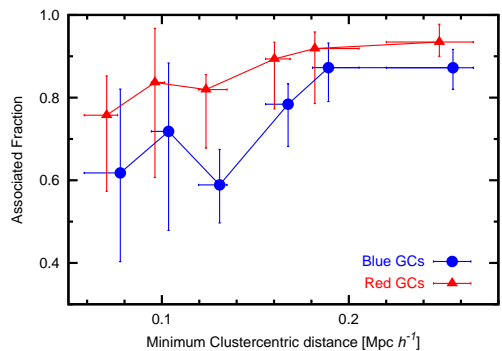}
\caption{Associated fraction of blue globular clusters at $z=0$ as a function of the minimum clustercentric distance. Points correspond
to median values and vertical error bars are obtained using the bootstrap resampling technique. Symbol and color types as in Figure \ref{fig:profiles}.}
\label{fig:fraccion_d}
\end{figure}

In Fig. \ref{fig:fraccion_orb}, we plot the fraction of globular clusters
at redshift $z=0$ as a function of the number of orbits; points correspond to 
median values and vertical error bars are computed using the bootstrap resampling technique. 
There is  a strong anticorrelation between the fraction of globular clusters
at redshift $z=0$ and the number of orbits. At least 30\%(25\%) of blue(red) globular clusters are stripped when halos have 
completed three or more orbits, while those that have completed only one orbit lose $\sim10\%$ of their globular
cluster populations. Based on their orbital trajectory, halos will cross the cluster center at 
different pericentric distances, experiencing
tidal forces of different strengths that modify the fraction of globular clusters 
assigned. Fig. \ref{fig:fraccion_d} shows the median fraction of globular clusters
associated with each halo at $z=0$ as a function of the minimun pericentric distance that each had throughout its whole
temporal evolution. Vertical error bars are computed using the 
bootstrap resampling technique in bins of pericentric distance containing equal numbers of halos. 
Those that during their lifetime approached the cluster core with 
pericentric distances less than
$150 \,\kpch$  lose approximately $\sim$40\%(25\%) of their blue(red) globular cluster population, 
while halos that remain away from the core with pericentric distances 
greater than $300 \,\kpch$ keep more than $\sim$80\%(90\%) of their blue(red) 
population. Both Fig. \ref{fig:fraccion_orb} and \ref{fig:fraccion_d} reinforce the
interpretation that blue globular clusters are more easily removed from halos 
than red ones due to the different slope of their density profiles. 
On average, at $z=0$, halos have lost $\sim 29\%$ of their initial blue globular clusters; this number drops to
$\sim 16\%$ for the red population, so it is expected that intracluster globular clusters will be mainly blue.

\begin{figure}
\includegraphics[width=7cm]{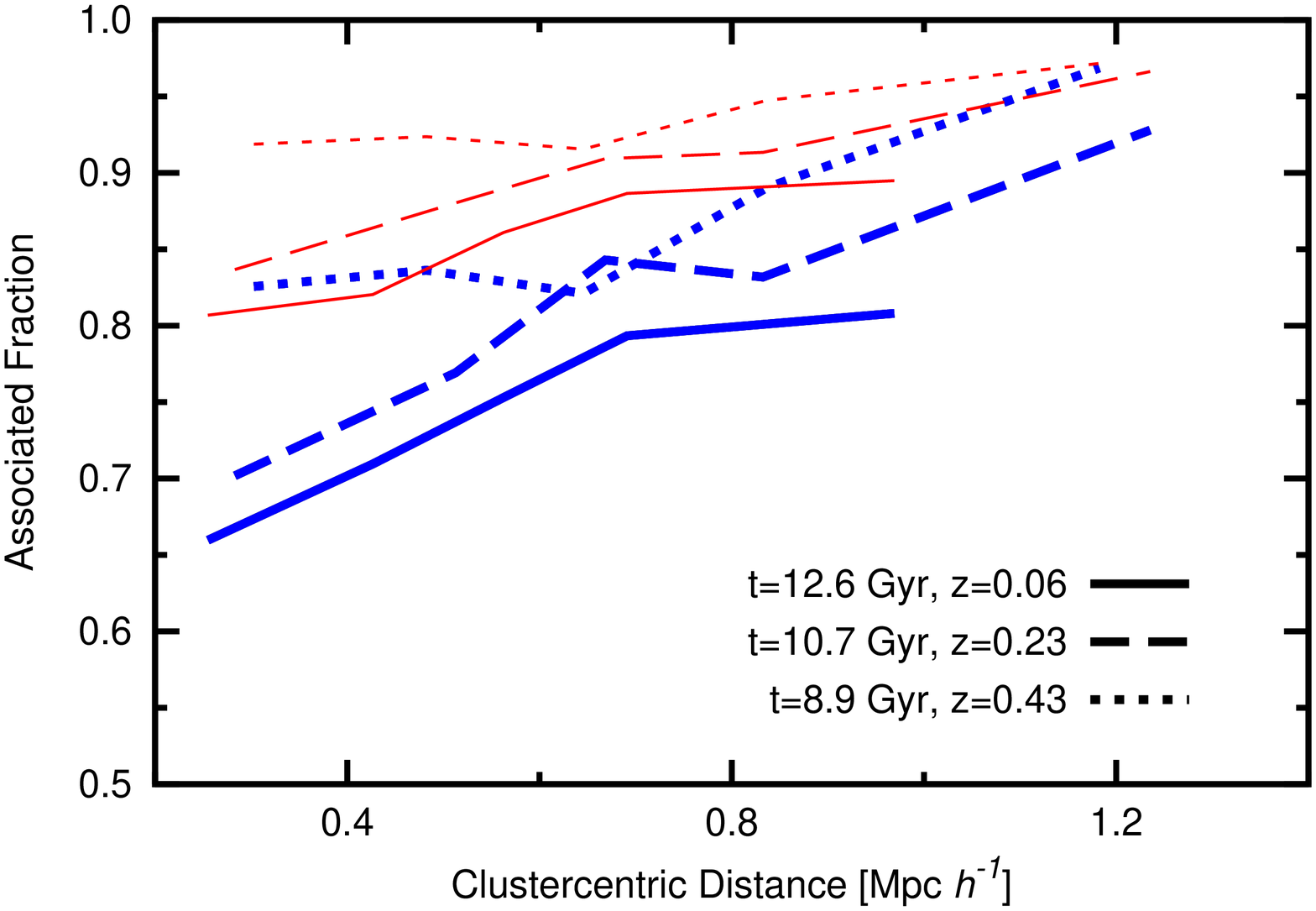}
  \caption{Median of the fraction of the blue and red globular clusters associated with its parent halo as a function of the clustercentric distance 
  at different redshifts: $z=0.06$ (solid line), $z=0.23$ (long dashed line) and $z=0.43$ (short dashed lines). 
  Color types as in Fig. \ref{fig:profiles}.}
\label{fig:median}
\end{figure}

The fraction of globular clusters that are associated with its halo at redshift $z=0$ shows a 
strong correlation with the number of orbits and the pericentric distance;
however, none of them are direct observables. 
In order to compare qualitatively our results with observations, we tested 
if the fraction of globular clusters associated, at any given redshift, is a
function of the clustercentric distance at that particular redshift. 
Fig. \ref{fig:median} shows the median of the fraction of globular clusters 
associated with their halo as a function of the clustercentric 
distance for both red and blue populations. Different line types correspond
to different redshift ranges, averaged as indicated by the labels.
We can see that associated clusters correlate with the clustercentric distance for 
three redshift ranges as shown. 
As expected, the effect depends on redshift, being stronger at redshift zero. 
Although the fraction of globular clusters that are associated with their halo is not a 
direct observable, these results can be compared with the findings of 
\citet{Coenda:2009} that 
the specific frequency ($S_N$) of the blue globular clusters in the Virgo Cluster increases with 
the clustercentric distance of galaxies.

\section{Discussion and conclusions}
\label{sec:conclusions}

Using a dark matter-only numerical simulation that 
follows the formation of a Virgo-like cluster of galaxies, we investigated the role of the 
cluster tidal field in the evolution of both red and blue globular cluster populations. At redshift $z = 0$, 
the simulated cluster has a virial mass $M_{vir} = 1.71 \times10^{14}\,M_{\odot}\, h^{-1}$, which is comparable to the mass of the Virgo Cluster.
We selected 38 dark matter halos with $M_{halo}\geq 1.08\times10^{10} M_{\odot} h^{-1}$ that entered the cluster virial radius when the 
Universe was younger than $10 \,\Gyr$. Using a Hernsquist profile, we randomly selected globular cluster particles 
in each dark matter halo, following the method outlined by \citet{Bullock:2005} and \citet{Pena:2008}. We used a 
Monte Carlo procedure to find the parameter that best reproduced the observed density profiles of both red and blue
globular clusters. We summarize our principle results as follows.

\begin{itemize}
 \item[-] Blue globular clusters are lost more easily than red ones. This is an expected result that may be explained by the shallower spatial 
distribution of blue globular clusters. The stripping of globular clusters is a continuous process that increases with the successive passages of 
halos through the center of the cluster. On average, halos have lost 16\% and 29\% of the red and blue globular cluster populations 
respectively, and therefore a significant number of globular clusters go to the intracluster medium.
 \item[-] None of the stripped globular clusters are recaptured by other halos.
 \item[-] Analyzing the bound fraction of globular clusters we found that at least $30\%$ of blue globular clusters are stripped when halos have orbited 
the cluster of galaxies three or more times. For those halos that have completed only one orbit the fraction removed
is $\sim10\%$.
 \item[-]We found a strong anticorrelation between the associated globular clusters and the minimum clustercentric distance that the halos have had throughout 
their history. Halos that have crossed the cluster core within $150\,\kpc$ from the cluster center during their lifetime, 
lost more than $40\%$ of their blue globular cluster population, while halos that remain away from the core with pericentric distances 
greater than $300\, \kpc$, have associated globular clusters greater than $80\%$. 
 \item[-] The associated globular clusters at a given time correlate with the clustercentric distance. The effect depends on redshift, being stronger at 
redshift zero. This result is in agreement with observational results \citep[see][]{Coenda:2009,Forbes:1997}.
  \item The density profile slope of \textbf{globular cluster systems} becomes slightly steeper immediately after the 
halos fall into the galaxy cluster. Thereafter, it remains almost constant, independently of the evolution of the 
host halo. This result suggests that the density profile of globular clusters remains almost equal over the lifetime 
of halos, in agreement with the ideas suggested by \citet{Coenda:2009}.   

\end{itemize}

Our results confirm that the stripping of globular clusters produced by the tidal field of clusters of galaxies is an efficient
process that may result in significant loss of the original globular cluster population. The magnitude of the effect will depend on 
the dynamical history of each galaxy. This phenomenon preferentially acts on the blue globular cluster population, and so it is expected 
that the intracluster globular clusters will be mainly blue.

\acknowledgments
We thank Aaron Ludlow and Julio Navarro for making these simulations available and for their comments, 
which improved an earlier version of this paper.
We acknowledge Rory Smith, Ken Freeman and Laura V. Sales’ useful suggestions.
We wish to thank the anonymous referee for useful suggestions that have improved the paper.
This work has been supported by grants from ANPCYT, CONICET and SECYT-UNC, Argentina.

\clearpage

\clearpage

\end{document}